\documentclass[12pt]{article}

\usepackage{epsf}
\usepackage{latexsym,amssymb,euscript}
\usepackage[dvips]{graphicx}
\usepackage{amsmath}

\begin{document}

\begin{center}
{\Large \textbf{The phase transitions in $2D$ $Z(N)$ vector models for $N>4$}}
\end{center}

\vskip 0.3cm
\centerline{O.~Borisenko$^{1\dagger}$, V.~Chelnokov$^{1*}$, 
G.~Cortese$^{2,3\P}$, R.~Fiore$^{3\P}$,
M.~Gravina$^{4\ddagger}$, A.~Papa$^{3\P}$}

\vskip 0.6cm
\centerline{${}^1$ \sl Bogolyubov Institute for Theoretical Physics,}
\centerline{\sl National Academy of Sciences of Ukraine,}
\centerline{\sl 03680 Kiev, Ukraine}

\vskip 0.2cm

\centerline{${}^2$ \sl Instituto de F\'{\i}sica Te\'orica UAM/CSIC,}
\centerline{\sl Cantoblanco, E-28049 Madrid, Spain}
\centerline{\sl and Departamento de F\'{\i}sica Te\'orica,}
\centerline{\sl Universidad de Zaragoza, E-50009 Zaragoza, Spain}

\vskip 0.2cm

\centerline{${}^3$ \sl Dipartimento di Fisica, Universit\`a della 
Calabria,}
\centerline{\sl and Istituto Nazionale di Fisica Nucleare, 
Gruppo collegato di Cosenza}
\centerline{\sl I-87036 Arcavacata di Rende, Cosenza, Italy}

\vskip 0.2cm

\centerline{${}^4$ \sl Department of Physics, University of Cyprus,
P.O. Box 20357, Nicosia, Cyprus}

\vskip 0.6cm

\begin{abstract}
We investigate both analytically and numerically the renormalization group 
equations in $2D$ $Z(N)$ vector models. The position of the critical points of 
the two phase transitions for $N>4$ is established and the critical index 
$\nu$ is computed.  
For $N$=7, 17 the critical points are located by Monte Carlo simulations
and some of the corresponding critical indices are determined. The behavior of 
the helicity modulus is studied for $N$=5, 7, 17. Using these and other 
available Monte Carlo data we discuss the scaling of the critical points with 
$N$ and some other open theoretical problems.   
\end{abstract}
 
\vfill
\hrule
\vspace{0.3cm}
{\it e-mail addresses}: 

$^\dagger$oleg@bitp.kiev.ua, \ \ $^*$vchelnokov@i.ua,
\ \ $^{\P}$cortese, fiore, papa \ @cs.infn.it, 

$^{\ddagger}$gravina@ucy.ac.cy

\newpage

\section{Introduction}
\label{intro}

The Berezinskii-Kosterlitz-Thouless (BKT) phase transition was originally 
discovered in the two-dimensional ($2D$) $XY$ model in the first half of 
the seventies~\cite{BKT,BKTRG}. 
Since then it was realized that this type of phase transition takes place in a 
number of other models including discrete $2D$ $Z(N)$ models for large enough 
$N$ and even $3D$ gauge models at finite temperature 
(see~\cite{3du1ft,3du1full} for a recent study of the deconfinement transition 
in $3D$ $U(1)$ lattice gauge theory). Here we are interested in the phase 
structure of $2D$ $Z(N)$ vector models~\footnote{See Ref.~\cite{noi} for
a shorter presentation of the results of this work.}. 
On a $2D$ lattice $\Lambda = L^2$ 
with linear extension $L$ and periodic boundary conditions, the partition 
function of the model can be written as
\begin{equation}
Z(\Lambda, \beta ) =\left[ \prod_{x\in \Lambda} 
\frac{1}{N} \sum_{s(x)=0}^{N-1} \right ]  \ \exp \left[ \sum_{x\in\Lambda} 
\ \sum_{n=1,2} \ \beta \ \cos\frac{2\pi }{N}( s(x)-s(x+e_n) ) \right]  \ .
\label{PFZNdef}
\end{equation}
The BKT transition is of infinite order and is characterized by the essential 
singularity, {\it i.e.} the exponential divergence of the correlation length. 
The low-temperature or BKT phase is a massless phase with a power-law decay of 
the two-point correlation function governed by a critical index $\eta$.  
The $Z(N)$ spin model in the Villain formulation has been studied analytically 
in Refs.~\cite{Elitzur,Villain}. It was shown that the model has at least two 
BKT-like phase transitions when $N\geq 5$.  The critical index $\eta$ has been 
estimated both from the renormalization group (RG) approach of the 
Kosterlitz-Thouless type and from the weak-coupling series for the
susceptibility. It turns out that $\eta(\beta^{(1)}_{\rm c})=1/4$ at the 
transition point from the strong coupling (high-temperature) phase to the 
massless phase, {\it i.e.} the behavior is similar to that of the $XY$ model. 
At the transition point $\beta^{(2)}_{\rm c}$ from the massless phase to 
the ordered low-temperature phase, one has $\eta(\beta^{(2)}_{\rm c})=4/N^2$. 
A rigorous proof that the BKT phase transition does take place, and so that the
massless phase exists, has been constructed in Ref.~\cite{rigbkt} for both 
Villain and standard formulations.
Monte Carlo simulations of the standard version with $N$=6, 8, 12 were 
performed in Ref.~\cite{cluster2d}. Results for the critical index $\eta$ 
agree well with the analytical predictions obtained from the Villain 
formulation of the model.
In Refs.~\cite{z5_lat10,z5_phys.rev} we have started a detailed numerical 
investigation of the BKT transition in $2D$ $Z(N)$ models for $N=5$, which is 
the lowest number where this transition can occur. Our findings support the 
scenario of two BKT transitions with conventional critical indices. 

Despite of this progress, some theoretical issues remain unresolved. 
\begin{itemize}
\item 
Status of the critical index $\nu$. This index governs the behavior of the 
correlation length $\xi$, namely it is expected that in the vicinity of the 
phase transition one has $\log\xi \sim a /(\beta_c-\beta)^{\nu}$. In all 
numerical studies one postulates the $XY$ value $\nu=1/2$. This value appears 
compatible with numerical data. 
However, no theoretical arguments exist which support this value. Moreover, 
the study of the Hamiltonian version of $Z(N)$ models via the strong coupling 
expansion combined with the Pad\'e approximant points to a different value 
for $N$=5, 6. Only for $N=12$ one finds $\nu\approx 1/2$~\cite{Elitzur}.  

\item
The scaling of critical points $\beta_c^{(1,2)}$ with $N$ is still unknown. 
What is seen from the available numerical data is that $\beta_c^{(1)}$ 
approaches the $XY$ critical point rather fast, probably exponentially with 
$N$, while $\beta_c^{(2)}$ scales like some power of $N$. 

\end{itemize}

Most theoretical knowledge about the critical behavior of the $XY$ model 
comes from the solution of the RG equations~\cite{BKTRG}. 
Explicit solutions for the RG equations of the arbitrary $Z(N)$ model are 
unknown. Moreover, to the best of our knowledge, even approximate and/or 
numerical solutions have not been obtained. 

One of the goals of the present paper is to fill this gap by constructing 
both numerical and approximate analytical solutions to the RG equations. 
Another goal is to continue the numerical study of $2D$ $Z(N)$ vector models 
and to compare results with analytical predictions.   

The paper is organized as follows. In the next section we investigate the 
system of RG equations describing the critical behavior of $Z(N)$ models in 
the Villain formulation. This allows us to calculate the position of critical 
points for all $N$ and to compute the critical index $\nu$. In the section~3 
we study numerically the models for $N$=7 and 17~\footnote{We chose
$N=7$ since this value interpolates $N=6$ and $N=8$ which were considered
in Ref.~\cite{cluster2d} and could be used as check-points; there is no special
reason for the choice of $N=17$, except that we made a balance between the 
need of a value considerably larger than the largest one considered so far in 
the literature ($N=12$) and a not too large one to slow down numerical 
simulations.}. 
We locate the transition 
points and compute some critical indices. Then we present and discuss our 
results for the helicity modulus. Section~4 is devoted to the analysis of the 
dependence of the critical points on $N$ and the comparison with the RG 
prediction. Finally, we summarize our results in section~5. 

\section{Analysis of RG equations}
\label{rg_eqn}

The system of RG equations describing the critical behavior of $2D$ $Z(N)$ 
models is given by~\cite{Elitzur}
\begin{equation}
\frac{dx}{d\tau} \ = \  \frac{N^2}{4} \ z^2 - x^2 y^2 \ \ , \;\;\;\;\;\;\;   
\frac{dy}{d\tau} \ = \  (2-x) y  \ \ , \;\;\;\;\;\;\;   
\frac{dz}{d\tau} \ = \ \left  (2-\frac{N^2}{4 x} \right ) z \ ,
\label{rgsysfull}
\end{equation}
where the parameters $x,y,z$ are initially defined as 
\begin{equation}
x=\pi\beta \ , \;\;\;\;\;\;\;
 y=2\pi\exp\left (-\frac{1}{2}\pi^2\beta \right ) \ , \;\;\;\;\;\;\; 
z=2\pi\exp \left (-\frac{N^2}{8\beta} \right ) \ ,
\label{initialval}
\end{equation}
and $a=\exp(\tau)$ is the lattice spacing~\footnote{We remind that there are 
two types of vortices in the Coulomb gas representation of the partition 
function of the $2D$ $Z(N)$ model residing on the direct and the dual 
lattices. Variables $y$ and $z$ can be related to the corresponding 
self-energies of these configurations.}. 

This system of RG equations has been obtained for the Villain formulation of 
the model. All available numerical data show that the standard and the 
Villain formulations belong to the same universality class in the case of the 
$XY$ model. This probably holds also for $Z(N)$ models. 
The only case where universality has been questioned is $Z(5)$, where it was 
found numerically that the helicity modulus has a different behavior in the 
vicinity of $\beta_c^{(1)}$: while it jumps to zero crossing the critical 
point in the Villain formulation, it goes to zero smoothly in the standard 
formulation~\cite{BM10}.      
 
\begin{figure}[tb]
\begin{center}
\includegraphics[width=0.48\textwidth]{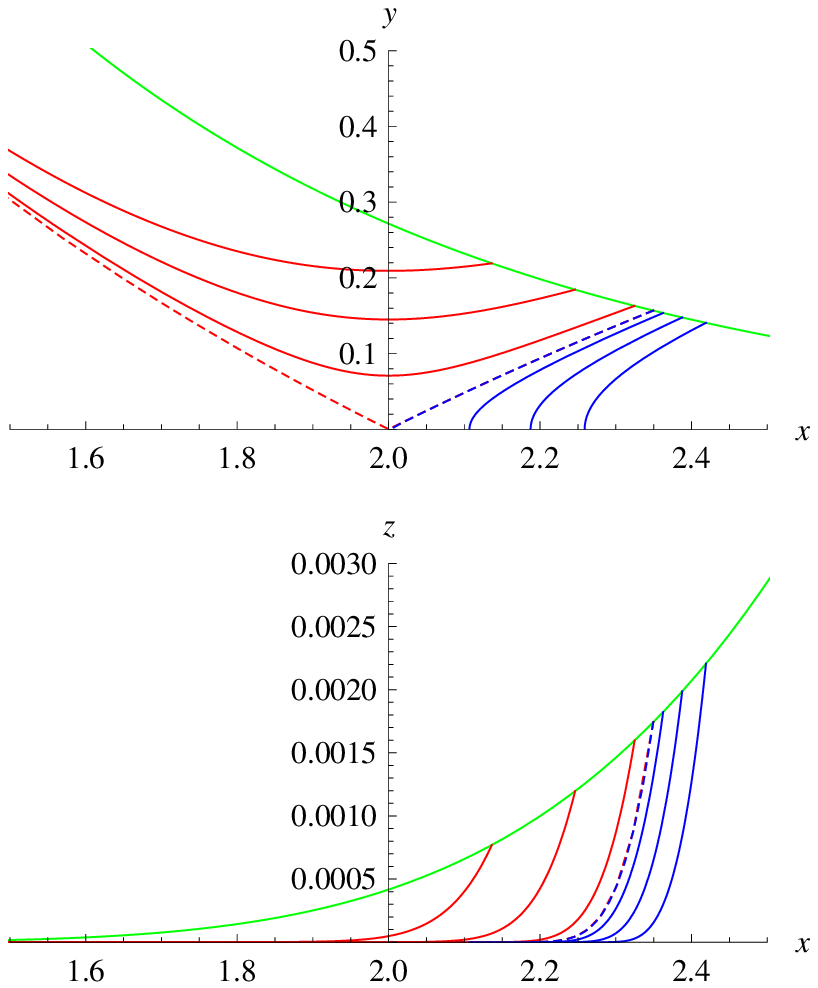}
\includegraphics[width=0.48\textwidth]{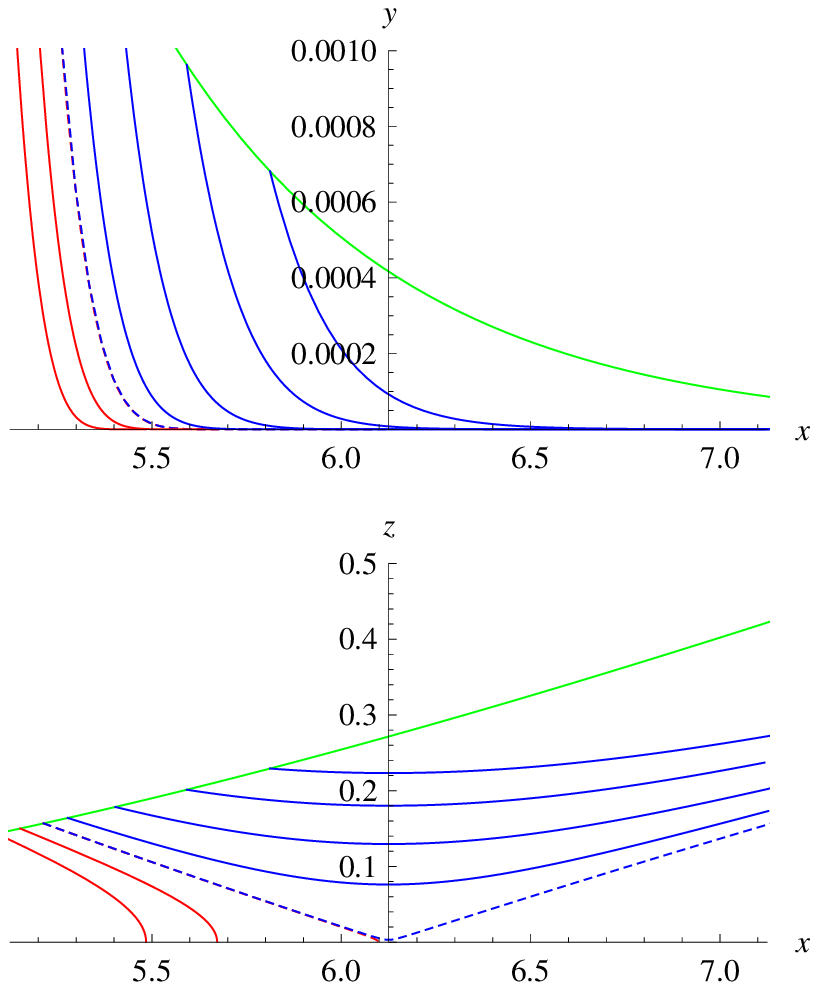}
\caption{(Color online) Renormalization group trajectories for $N=7$ in the 
vicinity of the first phase transition (left) and in the vicinity of the 
second phase transition (right). The upper line shows initial values.}
\label{fig:numericZN_2}
\end{center}
\end{figure}

\begin{table}[tb]
    \caption{Values of $\beta_c^{(1,2)}$ obtained from analytical and  
numerical solutions of Eqs.~(\ref{rgsysfull}) for various values of $N$.}
    \begin{center}
    \begin{tabular}{|c|c|c|c|c|}
    \hline
        $N$ & $\beta_{c \ \rm an}^{(1)}$ &  $\beta_{c \ \rm num}^{(1)}$ & 
$\beta_{c \ \rm an}^{(2)}$ &  $\beta_{c \ \rm num}^{(2)}$  \\
    \hline
    5	 & 0.726050 &  0.741654 & 0.841969 &  0.853845 \\
    6	 & 0.739670 &  0.747749 & 1.212435 &  1.219515 \\
    7	 & 0.740256 &  0.747851 & 1.650259 &  1.659667 \\
    8	 & 0.740266 &  0.747853 & 2.155441 &  2.167726 \\
    9	 & 0.740266 &  0.747853 & 2.727980 &  2.743528 \\
  $\infty$ & 0.7403 &  0.7479   &   -      &      -     \\ 
    \hline
    \end{tabular}
    \end{center}
    \label{tbl:betaci}
\end{table}

The system~(\ref{rgsysfull}) has two sets of the fixed points. The first one 
\begin{equation}
x=2 \ , \;\;\;\;\; y=0 \ , \;\;\;\;\; z=0 \ ,
\label{fixp1}
\end{equation}
corresponds to a transition from the massive to the massless phase. 
The second one 
\begin{equation}
x=\frac{N^2}{8} \ , \;\;\;\;\; y=0 \ , \;\;\;\;\; z=0 \ ,
\label{fixp2}
\end{equation}
describes the transition from the massless to the ordered phase. 
One can see that in the vicinity of the first fixed point the behavior of the 
solution in the $\tau\to\infty$ limit is the same as in the $XY$ model, 
{\it i.e} for $\beta\geqslant\beta_c^{(1)}$, $x$ tends to 
$x_\infty\geqslant2$, $y$ tends to $0$, and for $\beta<\beta_c^{(1)}$, $y$ 
tends to infinity ($z$ tends to zero in both cases).
Thus, $\beta_c^{(1)}$ may be determined as the value of $\beta$ at which 
the solution goes to the point $x=2,\ y=z=0$. Clearly, $\beta_c^{(2)}$ can be 
determined in a similar fashion, studying the system around the second fixed 
point $\beta = \frac{N^2}{8 \pi}$ (the behavior of $y$ and $z$ is reversed). 
Then, one can easily compute the RG trajectories starting from the initial 
values~(\ref{initialval}). As an example, RG trajectories, projected onto 
$x$-$y$ and $x$-$z$ planes, are shown in Fig.~\ref{fig:numericZN_2} for $N=7$. 
Dashed lines show the trajectory at the critical point. Numerical values of 
critical points for various values of $N$ are given in Table~\ref{tbl:betaci}.

To establish how the critical points scale with $N$ and to calculate the 
critical index $\nu$, it is desirable to have an analytical solution for the 
system~(\ref{rgsysfull}). In the absence of the exact solution, we attempt 
here to construct an approximate solution, valid for large values of $N$.
Making the following change of variables 
\begin{eqnarray} 
x \ = \ 2+\frac{1}{2} \ \xi \ , \;\;\;\;\;\;\;
 y \ = \ \sqrt{\frac{\omega}{8}} \ , \;\;\;\;\;\;\;
z \ = \ \exp \left [ \left ( 2-\frac{N^2}{8} \right ) \tau 
+  \frac{N^2}{32} \ \eta \right ]\;,
\label{newvar1}
\end{eqnarray}
and then linearizing the system~(\ref{rgsysfull}) around the first fixed point,
one finds 
\begin{equation}
\frac{d\xi}{d\tau} \ = \  - \omega + \frac{N^2}{2} \ \epsilon(\eta) \ \ ,  
\;\;\;\;\;\;\;
\frac{d\omega}{d\tau} \ = \  -\xi \ \omega  \ \ ,  \;\;\;\;\;\;\;
\frac{d\eta}{d\tau} \ = \ \xi \ .
\label{rgsys1}
\end{equation}
Here
\begin{equation} 
\epsilon(\eta) \ = \ \exp \left [ \left ( 4-\frac{N^2}{4} \right ) \tau 
+  \frac{N^2}{16} \ \eta \right ] \ = \ z^2
\label{epsilondef}
\end{equation}
is treated as a small function for large $N$. This is valid in two cases: 
1) $N\gg 4$ and 2) $\tau\to\infty$ if $N>4$. 

To study the system around the second fixed point $x=\frac{N^2}{8}$, we make 
the following change of variables: 
\begin{eqnarray} 
x \ = \ \frac{N^2}{8} \ (1-\xi/4) \ , \;\;\;\;\;\;\; 
y \ = \ \exp \left [ \left ( 2-\frac{N^2}{8} \right ) \tau 
+  \frac{N^2}{32} \ \eta \right ] \ , \;\;\;\;\;\;\; 
z \ = \ \sqrt{\frac{\omega}{8}} \ .
\label{newvar2}
\end{eqnarray}
The linearization of the system~(\ref{rgsysfull}) around the second fixed 
point leads to the same system of equations~(\ref{rgsys1}). This means that 
the solution of the system in terms of $\xi$, $\omega$ and $\epsilon$ will be 
the same as for the first critical point - we only have different initial 
conditions. 

To the first order in $\epsilon(\eta)$, one obtains 
\begin{equation} 
x \ = \ 2+ \frac{1}{2} \ (\xi_0 + \xi_1) \ , \;\;\;\;\;\;\; 
\omega \ = \ - \frac{d\xi_0}{d\tau} - \frac{d\xi_1}{d\tau} + 
\frac{N^2}{2} \ \epsilon(\eta_0) \ , \;\;\;\;\;\;\; 
\eta_0 \ = \ \int \xi_0 d\tau \ + \ C_3 \ . 
\label{xsol}
\end{equation}
Here $\xi_0$ is the solution of the system in the limit $N\to\infty$,
\begin{equation} 
\frac{d\xi_0}{d\tau} + \frac{1}{2} \ \xi_0^2 \ = \ C \ ,
\label{xi0eq}
\end{equation}
$\xi_1$ describes the first nontrivial corrections and obeys the following 
linear equation: 
\begin{equation} 
\frac{d\xi_1}{d\tau} + \xi_0  \xi_1 \ = \ g(\tau) \ , \;\;\;\;\;\;\; 
g(\tau) \ = \ \frac{N^2}{2} \ \epsilon(\eta_0)  + 
\frac{N^2}{2} \ \int \epsilon(\eta_0) \ \xi_0 \ d\tau \ .
\label{xi1eq}
\end{equation}
The solution for $\xi_0$ reads 
\begin{equation} 
\xi_0 \ = \ 2C_1 \coth{C_1(\tau-\tau_0)} \ , \;\;\;\;\;\;\;
C_1^2 \ = \ \frac{1}{2} \ C \ ,
\label{xi0sol}
\end{equation}
while the solution for $\xi_1$ vanishing for $N\to\infty$ can be presented as 
\begin{equation} 
\xi_1 \ = \ \exp(-\eta_0) \ \int g(\tau) \exp(\eta_0) \ d\tau \ .
\label{xi1sol}
\end{equation}
To find critical points, consider the zero-order approximation~(\ref{xi0sol}).
In the limit $C\to0$ (critical line of the model) and taking 
$C_3 = -2 \ln\sinh{C_1(C_2-\tau_0)}$, the solution can be written as 
\begin{equation}
\xi_0 \ = \ \frac{2}{\tau-\tau_0} \ , \;\;\;\;\;\;\; 
\omega_0 \ = \ \frac{2}{\left(\tau-\tau_0\right)^2} \ , \;\;\;\;\;\;\;
\eta_0 \ = \ 2 \ln \frac{\tau-\tau_0}{C_2-\tau_0} \ .
\label{xi0solcrit}
\end{equation}
We fix $\tau_0$ from the requirement that $\xi_0(0)$ and $\omega_0(0)$ 
satisfy the initial conditions for $x$ and $y$ given in~(\ref{initialval}) 
(this effectively means that $\xi_0$ and $\omega_0$ describe the critical 
trajectory of the $XY$ model). Hence, 
\begin{equation}
2 \pi \exp{ \left(-\frac{1}{2} \pi \left( 2 -\frac{1}{\tau_0} \right)\right)} 
\ = \ - \frac{1}{2 \tau_0} \ , \;\;\;\;\;\;\;
 \ \tau_0 = \frac{1}{2 - \pi \beta_c^{XY}} \ .
\label{tau0eq}
\end{equation}
From the last equation one finds $\beta_c^{XY} \approx 0.7403$. 
It is now straightforward to calculate first order corrections to this 
solution. After a long algebra one finds the following equation: 
\begin{equation}
32 \pi^2 \ \exp{ \left [ - \pi^2 \beta_c^{(1)} \right ] }\ = \ 
\frac{2}{\tau_0^2} - \frac{4 \pi}{\tau_0} \left ( \beta_c^{(1)} - \beta_c^{XY} 
\right ) +  2 \pi^2  N^2 \ \exp{ \left [ - \frac{N^2}{4 \beta_c^{(1)}} 
\right ] } \ .
\label{betac1eqs}
\end{equation}
An approximate solution is given by 
\begin{equation}
\beta_c^{(1)} \ = \ \beta_c^{XY} - 
\frac{\pi N^2}{\left ( \pi \beta_c^{XY} - 2 \right ) \left ( 2 - 2 \pi 
+ \beta_c^{XY} \pi^2 \right )} \ 
\exp{ \left [ - \frac{N^2}{4 \beta_c^{XY}} \right ] } \ .
\label{betac1appr}
\end{equation}

The same strategy applied for the second fixed point leads to the equation 
for $\beta_c^{(2)}$:
\begin{eqnarray}
\beta_c^{(2)} \ = \ 
\frac{N^2}{8 \pi} \left ( 1 + \frac{1}{2 \tau_0} \right ) - 
\frac{1}{2} \pi^3 \tau_0 N^2 \ \exp{ \left [ - \frac{\pi N^2}{8} 
\left ( 1 + \frac{1}{2 \tau_0} \right )\right ] } \ ,
\label{betac2approx}
\end{eqnarray}
where $\tau_0$ is a solution of the equation
\begin{equation}
\exp{ \left [ - \frac{2 \pi \tau_0}{2 \tau_0 + 1} \right ] } \ 
= \ -\frac{1}{4 \pi \tau_0} \ .
\label{tau0betac20}
\end{equation}

These results are summarized in the Table~\ref{tbl:betaci}: 
$\beta_{c \ \rm an}^{(1,2)}$ is the approximate value computed 
from~(\ref{betac1eqs}) (first critical point) and 
from~(\ref{betac2approx})-(\ref{tau0betac20}) (second critical point); 
$\beta_{c \ \rm num}^{(1,2)}$ is the numerical value computed from the 
system~(\ref{rgsysfull}).
The last row gives the critical values for the $XY$ model: the analytical 
solution $\beta_c = 0.7403$ of the linearized system and the numerical 
solution $\beta_c = 0.7479$ of the exact equations~(\ref{rgsysfull}). 
This RG prediction should be compared with the Monte Carlo result 
$\beta_c \approx 0.751$ for the Villain model~\cite{XYVillaincr}.

Finally, we would like to gain some information on the critical index $\nu$. 
To compute $\nu$ one has to construct the solution in the region 
$y\to\infty$ for $\tau\to\infty$ (for the first transition) and 
$z\to\infty$ for $\tau\to\infty$ (for the second transition). 
These are the two regions with non-vanishing mass gap. 
The corresponding solutions can be easily derived from the solutions 
described above. It is clear, however, that by the very virtue 
of that construction, the leading singularity will be the same as 
in the $XY$ model. This means $\nu=1/2$ for all large enough $N$. 
Therefore, it is much more informative if we consider models with 
$N$ not too large and construct fits for $\tau$ from the numerical solutions 
of the RG equations~(\ref{rgsysfull}) in the given regions.    
We remind that 
\begin{equation}
\tau \ = \ \log a \ \sim \ \log\xi \ ,
\label{xi}
\end{equation}
where $\xi$ is the correlation length. 
Hence, we fit $\tau$ with the following function 
\begin{equation}
\tau = A + \frac{B}{(\beta_c-\beta)^{\nu}} \ .
\label{fit_nu} 
\end{equation}
In Tables~\ref{tbl:nu_1} and~\ref{tbl:nu_2} we give results for $N$=5,
6, 7, 8, 9. The number of fitting points together with the maximal value 
$(\beta_c-\beta)$ are also shown in the Tables. Fitting points have been 
distributed uniformly between $\beta_c-\beta = 0$ and $\max{(\beta_c-\beta)}$. 
Only central values for all coefficients in both Tables are shown.

\begin{table}[tb]
    \caption{Index $\nu$ derived from~(\ref{rgsysfull}) and~(\ref{fit_nu}) 
for the first phase transition.}
    \begin{center}
    \begin{tabular}{|c|c|c|c|c|c|c|c|}
    \hline
$N$ & $A$ & $B$ & $\nu$ & $\beta_c$ & $\chi^2$ & points & 
$\max{(\beta_c-\beta)}$\\
    \hline
5 & $-$1.78073 & 0.76176  & 0.516378 & 0.741654 & 2.5613     & 150 & 0.01 \\
5 & $-$2.77757 & 0.834452 & 0.507893 & 0.741654 & 0.00627118 & 100 & 0.005 \\
6 & $-$3.03443 & 0.914498 & 0.506457 & 0.747749 & 0.00744256 & 100 & 0.01 \\
7 & $-$3.09949 & 0.928961 & 0.504642 & 0.747851 & 0.00454228 & 100 & 0.01 \\
8 & $-$3.01409 & 0.910821 & 0.507149 & 0.747853 & 0.00435793 & 100 & 0.01 \\
9 & $-$3.01271 & 0.910531 & 0.507189 & 0.747853 & 0.00422122 & 100 & 0.01 \\
      \hline
    \end{tabular}
    \end{center}
    \label{tbl:nu_1}
\end{table}

\begin{table}[tb]
    \caption{Index $\nu$ derived from~(\ref{rgsysfull}) and~(\ref{fit_nu}) 
for the second phase transition.}
    \begin{center}
    \begin{tabular}{|c|c|c|c|c|c|c|c|}
    \hline
$N$ & $A$ & $B$ & $\nu$ & $\beta_c$ & $\chi^2$ & points & 
$\max{(\beta-\beta_c)}$\\
    \hline
5 & $-$2.56099 & 0.874616 & 0.510268 & 0.853845 & 2.12609 & 150 & 0.01\\
5 & $-$2.65193 & 0.888885  & 0.508654 & 0.853845 & 0.0243011 & 100 & 0.005 \\
6 & $-$3.07805 & 1.18864  & 0.504577 & 1.21951  & 0.0212314 & 100 & 0.01 \\
7 & $-$3.04924 & 1.38005  & 0.505627 & 1.65967  & 0.00454228 & 100 & 0.01 \\
8 & $-$3.22569 & 1.61538  & 0.502838 & 2.16773  & 0.0271673 & 100 & 0.01 \\
9 & $-$3.25341 & 1.82077  & 0.502766 & 2.74353  & 0.0212578 & 100 & 0.01 \\
      \hline
    \end{tabular}
    \end{center}
    \label{tbl:nu_2}
\end{table}

Deviations from the central values are in general very small, except for the 
coefficient $A$ for $N=5$ and when $\max{(\beta_c-\beta)}=0.01$. The situation 
is much improved if one takes $\max{(\beta_c-\beta)}=0.005$. This can indicate 
that the scaling region in $Z(5)$ model is somewhat narrower than for $N>5$. 
We have also tried to fit $\tau$ assuming the power-like singularity for the 
correlation length. The quality of fits is poor in this case and 
$\nu$ acquires a very large value which varies with varying $N$. This leaves 
no doubts that the correlation length diverges exponentially with $\nu=1/2$ 
for all $N>4$ in the vicinity of both phase transitions. 


\section{Numerical results}
\label{numerical}

We simulated the model defined by Eq.~(\ref{PFZNdef}) using the same cluster 
Monte Carlo algorithm adopted in the Refs.~\cite{z5_lat10,z5_phys.rev} for the 
case $N=5$.
We used several different observables to probe the two expected phase 
transitions. In order to detect the first transition ({\it i.e.} the one
from the disordered to the massless phase) we used the absolute value 
$|M_{L}|$ of the complex magnetization,
\begin{equation}
M_{L}=\frac{1}{L^{2}}\sum_{i}\exp\left(i\frac{2\pi}{N}s_{i}\right)\equiv |M_{L}|e^{i\psi},
\label{magn_complex}
\end{equation}
and the {\em helicity modulus}~\cite{hel_def,torsion} 
\begin{equation}
\Upsilon=\left<e\right> - L^{2}\beta\left<s^{2}\right>,
\label{Helicity_formula}
\end{equation}
where $e\equiv\frac{1}{L^{2}} \sum_{<ij>_{x}} \cos\left(\theta_{i}-\theta_{j}
\right)$ and $s\equiv\frac{1}{L^{2}} \sum_{<ij>_{x}} \sin(\theta_{i}
-\theta_{j})$ and the notation $<ij>_{x}$ means nearest-neighbors spins
in the $x$-direction.

For the second transition ({\it i.e.} the one from the massless to the ordered 
phase) we adopted the real part of the "rotated" magnetization, 
\[
M_{R}=|M_{L}|\cos(N\psi)\;,
\]
and the order parameter
\[
m_{\psi}=\cos(N\psi)
\]
introduced in Ref.~\cite{BMK09}, where $\psi$ is the phase of the complex 
magnetization defined in Eq.~(\ref{magn_complex}). 
In this work, both for $N=7$ and $N=17$, we 
collected typically 100k measurements for each value of the coupling $\beta$, 
with 10 updating sweeps between each configuration. To ensure  thermalization 
we discarded for each run the first 10k configurations. The jackknife method 
over bins at different blocking levels was used for the data analysis.

\begin{figure}[tb]
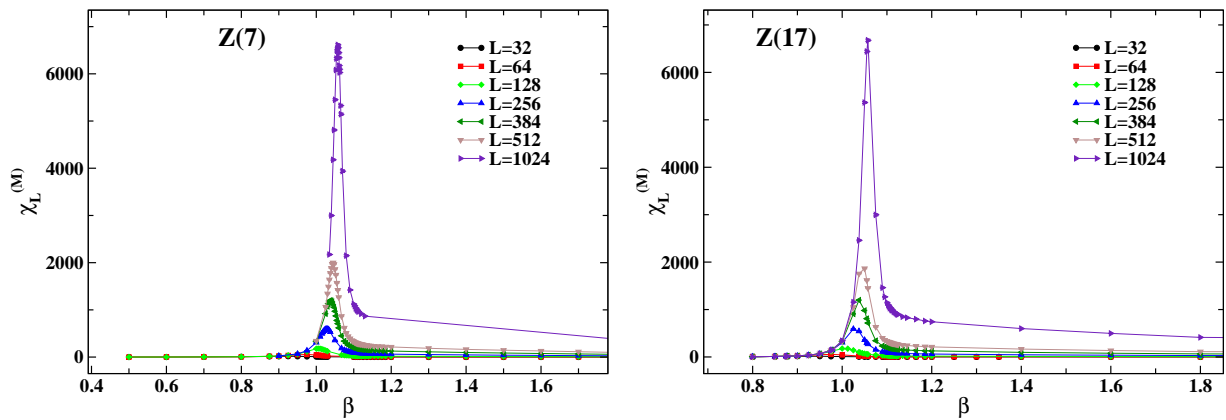

\begin{center}
\includegraphics[scale=0.32]{magn_real_suscet_q7.eps}\hspace{0.2cm}
\includegraphics[scale=0.32]{magn_real_suscet_q17.eps}
\caption{(Color online) Susceptibility $\chi_{L}^{(M)}$ versus $\beta$ 
in $Z(7)$ (left) and $Z(17)$ (right) on lattices with several values of $L$.}
\label{Magn_suscet}
\end{center}
\end{figure}  

\begin{figure}[tb]
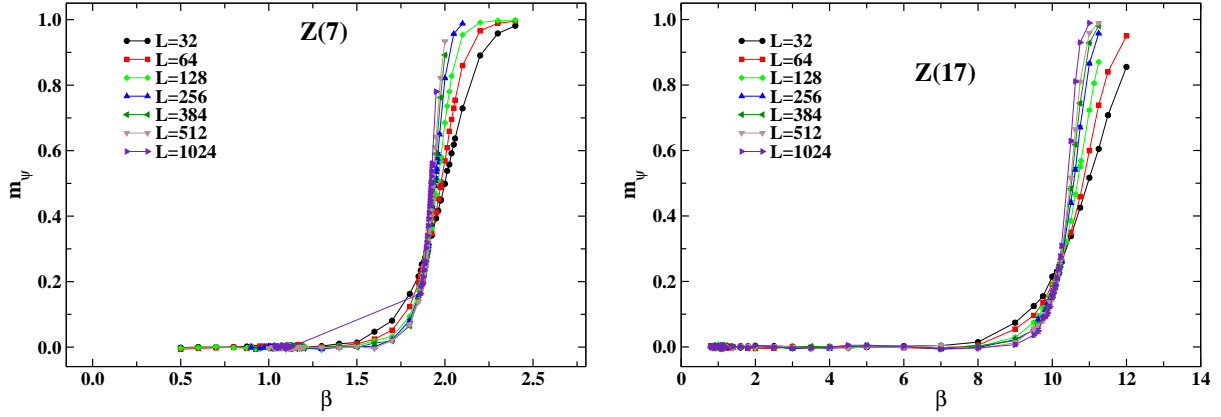

\begin{center}
\includegraphics[scale=0.32]{mpsi_q7.eps}
\hspace{0.2cm}
\includegraphics[scale=0.32]{mpsi_q17.eps}
\end{center}
\vspace{-0.5cm}
\caption{(Color online) Behavior of $m_{\psi}$ with $\beta$ in $Z(7)$ (left) 
and $Z(17)$ (right) on lattices with several values of $L$.}
\label{Mpsi}
\end{figure}

In Fig.~\ref{Magn_suscet} we show the behavior of the susceptibility 
$\chi_{L}^{(M)}\equiv L^2 (\langle |M_L|^2\rangle - \langle |M_L|\rangle^2)$
of the absolute value of the complex magnetization, which
exhibits, for each volume considered, a clear peak signalling the first phase
transition. The position of the peak in the thermodynamic limit defines 
the first critical coupling, $\beta_{c}^{(1)}$. Fig.~\ref{Mpsi} shows instead 
the behavior of $m_\psi$ versus $\beta$ on various lattice sizes; here the 
second critical coupling $\beta_{c}^{(2)}$ is identified by the crossing 
point (in the thermodynamic limit) of the curves formed by the data on
different lattice sizes. 

To determine the first critical coupling $\beta_{c}^{(1)}$, we could 
extrapolate to infinite volume the pseudo-critical couplings 
given by the position of the peaks of $\chi_{L}^{(M)}$. However, since the 
approach to the thermodynamic limit is rather slow (powers  of $\log L$),
we adopted a different method, based on the use of the 
``\textsl{reduced fourth-order}'' Binder cumulant 
\begin{equation}
U^{(M)}_L=1-\frac{\langle |M_L|^4 \rangle}{3\langle |M_L|^2 \rangle^2} \; ,
\label{binder_U}
\end{equation}
the cumulant $B_4^{(M_R)}$ defined as
\begin{equation}
B_4^{(M_R)}=\frac{\langle |M_R-\langle M_R\rangle|^4\rangle}
{\langle |M_R-\langle M_R\rangle|^2\rangle^2}\;,
\label{binder_MR}
\end{equation}
and the helicity modulus $\Upsilon$.
We estimated $\beta_c^{(1)}$ by looking for (i) the crossing point of the 
curves, obtained on different volumes, giving a Binder cumulant versus 
$\beta$ and (ii) the optimal overlap of the same curves after 
plotting them versus $(\beta - \beta_{c}) (\log L )^{1/\nu}$, with $\nu$ fixed
at 1/2. The method (ii) has been applied also to the helicity modulus 
$\Upsilon$. Our best values for $\beta_{c}^{(1)}$ are
\begin{eqnarray*}
N=7:&& \;\;\; \beta_{c}^{(1)}=1.1113(13) \;, \\
N=17:&&  \;\;\; \beta_{c}^{(1)}=1.11375(250)\;.
\end{eqnarray*}
Then, we performed the finite size scaling (FSS) analysis of the magnetization 
$|M_{L}|$ and the susceptibility $\chi_{L}^{(M)}$ at $\beta_{c}^{(1)}$
using the following laws:
\begin{equation}
|M_{L}|(\beta_{c}^{(1)}) = A L^{-\beta/\nu} \;, 
\label{law_magn}
\end{equation}
\begin{equation}
\chi_{L}^{(M)}(\beta_{c}^{(1)}) = B L^{\gamma/\nu}\;,
\label{law_susc}
\end{equation}
where $\gamma/\nu=2-\eta$ and $\eta$ is the {\em magnetic critical index}. 
Results are summarized in Tables~\ref{scaling_magn_7}, \ref{scaling_magn_17},  
\ref{scaling_suscet_magn_7} and \ref{scaling_suscet_magn_17}. 
We observe that the hyperscaling relation 
$\gamma/\nu + 2\beta/\nu = d=2$ is nicely satisfied within statistical errors 
in both models. 

\begin{table}[tb]
\centering
\caption[]{Results of the fit to the data of $|M_{L}|(\beta_{c}^{(1)})$ 
with the scaling law~(\ref{law_magn}) on $L^{2}$ lattices with 
$L\geqslant L_{\rm min}$, for $N=7$.}
\vspace{0.2cm}
\begin{tabular}{|c|c|c|c|}
\hline
\multicolumn{4}{|c|}{$N=7$}\\
\hline
 $L_{\rm min}$ & $A$ & $\beta/\nu$ & $\chi^{2}$/d.o.f. \\
\hline
 32  & 1.00653(48) & 0.12210(08) & 5.5  \\
 64  & 1.00858(70) & 0.12243(12) & 3.7  \\
 128 & 1.01074(94) & 0.12277(15) & 2.0  \\
 256 & 1.0146(16)  & 0.12336(26) & 0.40 \\
 384 & 1.0162(22)  & 0.12359(34) & 0.19 \\
 512 & 1.0177(38)  & 0.12381(56) & 0.16 \\
 640 & 1.0185(57)  & 0.12393(84) & 0.28 \\
\hline
\end{tabular}
\label{scaling_magn_7}
\end{table}

\begin{table}[tb]
\centering
\caption[]{Results of the fit to the data of $|M_{L}|(\beta_{c}^{(1)})$ 
with the scaling law~(\ref{law_magn}) on $L^{2}$ lattices with 
$L\geqslant L_{\rm min}$, for $N=17$.}
\vspace{0.2cm}
\begin{tabular}{|c|c|c|c|}
\hline
\multicolumn{4}{|c|}{$N=17$}\\
\hline
 $L_{\rm min}$ & $A$ & $\beta/\nu$ & $\chi^{2}$/d.o.f. \\
\hline	
  32 & 1.00388(51) & 0.12111(09) & 7.98 \\
  64 & 1.00620(69) & 0.12149(12) & 3.58 \\
 128 & 1.0089(11)  & 0.12191(18) & 1.54 \\
 256 & 1.0107(15)  & 0.12219(24) & 0.74 \\
 384 & 1.0113(24)  & 0.12228(36) & 1.36 \\
\hline
\end{tabular}
\label{scaling_magn_17}
\end{table}


\begin{table}[tb]
\centering
\caption[]{Results of the fit to the data of $\chi_{L}^{(M)}(\beta_{c}^{(1)})$ 
with the scaling law~(\ref{law_susc}) on $L^{2}$ lattices with 
$L\geqslant L_{\rm min}$, for $N=7$.}
\vspace{0.2cm}
\begin{tabular}{|c|c|c|c|}
\hline
\multicolumn{4}{|c|}{$N=7$}\\
\hline
 $L_{\rm min}$ & $B$ & $\gamma/\nu$ & $\chi^{2}$/d.o.f. \\
\hline
  32 & 0.00558(07) & 1.7402(23) & 2.38   \\
  64 & 0.00540(09) & 1.7457(29) & 1.32   \\
 128 & 0.00522(12) & 1.7508(38) & 0.68   \\
 256 & 0.00518(20) & 1.7520(61) & 0.84   \\
 384 & 0.00546(33) & 1.7443(93) & 0.70   \\	
 512 & 0.00489(52) & 1.760(16)  & 0.28   \\
 640 & 0.00444(76) & 1.775(25)  & 0.0066 \\

\hline
\end{tabular}
\label{scaling_suscet_magn_7}
\end{table}

\begin{table}[tb]
\centering
\caption[]{Results of the fit to the data of $\chi_{L}^{(M)}(\beta_{c}^{(1)})$ 
with the scaling law~(\ref{law_susc}) on $L^{2}$ lattices with 
$L\geqslant L_{\rm min}$, for $N=17$.}
\vspace{0.2cm}
\begin{tabular}{|c|c|c|c|}
\hline
\multicolumn{4}{|c|}{$N=17$}\\
\hline
 $L_{\rm min}$ & $B$ & $\gamma/\nu$ & $\chi^{2}$/d.o.f. \\
\hline	
  32 & 0.00559(08) & 1.7372(26) & 2.7  \\
  64 & 0.00532(11) & 1.7453(35) & 0.39 \\
 128 & 0.00521(15) & 1.7484(46) & 0.16 \\
 256 & 0.00514(23) & 1.7507(71) & 0.15 \\
 384 & 0.00522(31) & 1.7483(92) & 0.13 \\
\hline
\end{tabular}
\label{scaling_suscet_magn_17}
\vspace{1cm}
\end{table}

\begin{figure}[tb]
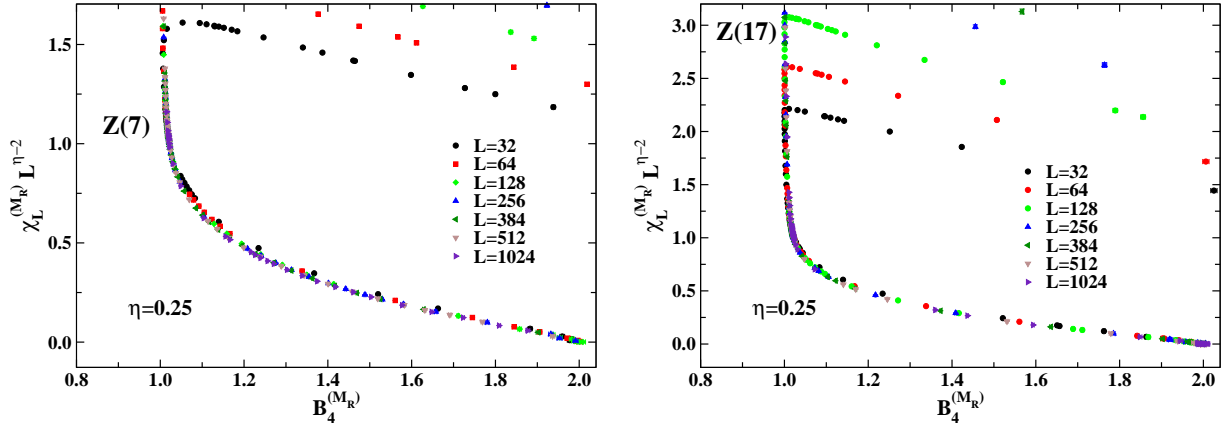

\begin{center}
\includegraphics[scale=0.32]{Susc_rot_vs_b4_eta=025.eps}
\hspace{0.2cm}
\includegraphics[scale=0.32]{Xi_rot_vs_binder_eta=025.eps}
\end{center}
\vspace{-0.5cm}
\caption{(Color online) Correlation between $\chi_{L}^{(M_{R})}L^{\eta-2}$ and 
the Binder cumulant $B_{4}^{(M_{R})}$ for $\eta=0.25$ in $Z(7)$ (left) and 
$Z(17)$ (right) on lattices with $L$ ranging from 128 to 1024.}
\label{Susc_rot_b4}
\end{figure}

We can cross-check our determination of the critical exponent $\eta$ by
an independent method, which does not rely on the prior knowledge of
the critical coupling, but is based on the construction of a suitable 
universal quantity~\cite{Loison,z5_phys.rev}. The idea is to plot 
$\chi_{L}^{(M_{R})}L^{\eta-2}$ versus $B_{4}^{(M_{R})}$ and to look for the 
value of $\eta$ which optimizes the overlap of curves from different volumes.
We found that, both in $Z(7)$ and $Z(17)$, $\eta=1/4$ is this optimal
value, since it gives the best overlap of these curves in the region of 
values corresponding to the first phase transition, {\it i.e.}
the lower branch of the curves of Fig.~\ref{Susc_rot_b4}. This result for 
$\eta$ agrees with the determinations $\eta=2-\gamma/\nu$ from the FSS 
analysis.

As for the second critical coupling $\beta_{c}^{(2)}$, we used the same method
adopted for $\beta_{c}^{(1)}$, but applied now to $B_{4}^{(M_{R})}$ 
and $m_{\psi}$. Our best estimates are
\[
N=7: \;\;\; \beta_{c}^{(2)}=1.8775(75)\;,
\;\;\;\;\;\;\;\;\;\;\;\;\;\;\;
N=17: \;\;\; \beta_{c}^{(2)}=10.13(12)\;.
\]
The standard FSS analysis applied to the susceptibility $\chi_{L}^{(M_{R})}$ 
of the rotated magnetization $M_{R}$ at $\beta_c^{(2)}$ leads to the result
for the critical indices $\gamma/\nu$ given in 
Tables~\ref{scaling_magn2_7} and~\ref{scaling_magn2_17}~\footnote{We do not 
report in this work the determinations of $\beta/\nu$ by the FSS analysis of 
the rotated magnetization $M_{R}$, since they are affected by large 
statistical and systematic uncertainties.}.

\begin{figure}[tb]
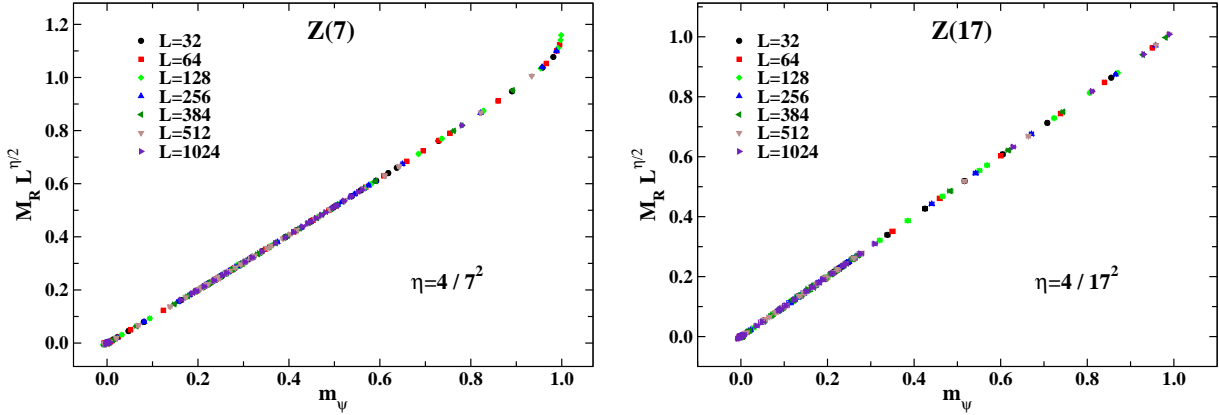

\begin{center}
\includegraphics[scale=0.32]{M_r_vs_mpsi_eta=0082.eps}
\hspace{0.2cm}
\includegraphics[scale=0.32]{M_vs_mpsi_eta=0014.eps}
\end{center}
\vspace{-0.5cm}
\caption{(Color online) Correlation between $M_{R}L^{\eta/2}$ and $m_{\psi}$ 
for $\eta=4/7^{2}$ in $Z(7)$ (left) and $\eta=4/17^{2}$ in $Z(17)$ (right), on 
lattices with L ranging from 128 to 1024.}
\label{M_mpsi}
\end{figure}  

\begin{table}[tb]
\centering
\caption[]{Results of the fit to the data of 
$\chi_{L}^{(M_{R})}(\beta_{c}^{(2)})$ with the scaling 
law~(\ref{law_susc})
on $L^{2}$ lattices with $L\geqslant L_{min}$, for $N=7$.}
\vspace{0.2cm}
\begin{tabular}{|c|c|c|c|}
\hline
\multicolumn{4}{|c|}{$N=7$}\\
\hline
$L_{min}$ & $A$ & $\gamma/\nu$ & $\chi^2$/d.o.f. \\
\hline	
  32 & 0.8767(37) & 1.92340(71) & 2.02 \\
  64 & 0.8833(47) & 1.92219(87) & 1.41 \\
 128 & 0.8858(57) & 1.9217(11)  & 1.57 \\
 256 & 0.8997(93) & 1.9193(16)  & 1.02 \\
 384 & 0.916(15)  & 1.9166(25)  & 0.68 \\
 512 & 0.921(24)  & 1.9158(39)  & 0.98 \\
 640 & 0.942(34)  & 1.9124(54)  & 1.07 \\
\hline
\end{tabular}
\label{scaling_magn2_7}
\end{table}

\begin{table}[tb]
\centering
\caption[]{Results of the fit to the data of 
$\chi_{L}^{(M_{R})}(\beta_{c}^{(2)})$ with the scaling 
law~(\ref{law_susc})
on $L^{2}$ lattices with $L\geqslant L_{min}$, for $N=17$.}
\begin{tabular}{|c|c|c|c|}
\hline
\multicolumn{4}{|c|}{$N=17$}\\
\hline
$L_{min}$ & $B$ & $\gamma/\nu$ & $\chi^2$/d.o.f. \\
\hline
  32 & 0.9319(47)  & 1.98933(89) & 1.65 \\
  64 &  0.9408(68) & 1.9878(12)  & 1.38 \\
 128 &  0.9533(89) & 1.9857(16)  & 0.67 \\
 256 &  0.954(16)  & 1.9856(27)  & 0.83 \\
 384 &  0.950(24)  & 1.9861(40)  & 1.09 \\	
 512 & 0.931(38)   & 1.9892(62)  & 1.44 \\
 640 & 0.911(59)   & 1.9925(98)  & 2.67 \\
\hline
\end{tabular}
\label{scaling_magn2_17}
\end{table}


\begin{figure}[tb]
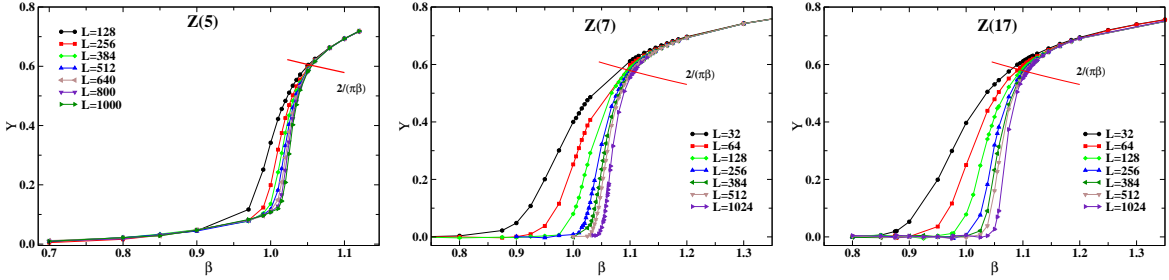

\vspace{0.5cm}
\begin{center}
\includegraphics[scale=0.21]{helicity_q5.eps}\hspace{0.1cm}
\includegraphics[scale=0.21]{helicity_q7.eps}\hspace{0.1cm}
\includegraphics[scale=0.21]{helicity_q17.eps}
\end{center}
\vspace{-0.5cm}
\caption{(Color online) Helicity modulus versus $\beta$ in $Z(5)$, $Z(7)$ and 
$Z(17)$ on lattices with various sizes.}
\label{Helicity}
\end{figure}  

Also in this case the critical index $\eta$ can be determined by an 
independent method, irrespectively of the knowledge of $\beta_c^{(2)}$:
$M_{R}L^{\eta/2}$ is plotted versus $m_{\psi}$ and the value of $\eta$ is
searched for, which optimizes the overlap of data points coming 
from different volumes. The results we found for $\eta$ in $Z(7)$ and $Z(17)$ 
are in perfect agreement with the theoretical prediction $\eta^{(2)}=4/N^{2}$
(see Fig.~\ref{M_mpsi}).

Finally, in Fig.~\ref{Helicity} we present the behavior with $\beta$ of the 
helicity modulus~(\ref{Helicity_formula}). This quantity is constructed in 
such a way that it should exhibit a discontinuous jump (in the thermodynamic 
limit) at the critical temperature separating the disordered phase from the 
massless one, if the transition is of infinite order (BKT). 
Since the Kosterlitz-Thouless RG equations for the $XY$ 
model~\cite{BKT,Ohta,Nelson} 
lead to the prediction that the helicity modulus $\Upsilon$ jumps from the 
value $2/(\pi \beta)$ to zero at the critical temperature, one can check 
if the same occurs for vector Potts models. In Fig.~\ref{Helicity} we plot a 
red line, representing the function $2/(\pi \beta)$; the crossing between this
line and the curves formed by data points of $\Upsilon$ approaches indeed
$\beta_{c}^{(1)}$ when the lattice size increases.

The knowledge of the behavior with $\beta$ of the helicity modulus
provides us with another method for the determination of the critical
index $\eta$. As shown in Ref.~\cite{Nelson,Him84} (see also Ref.~\cite{YO91}),
the following relation holds,
\begin{equation}
\eta = \frac{1}{2\pi\beta \Upsilon}\;,
\label{eta_vs_Y}
\end{equation}
which allows us to get the value of $\eta$ at any fixed $\beta$ in the 
BKT phase, if the value of $\Upsilon$ at that $\beta$ is known. A simple 
inspection of the behavior of $\Upsilon$ with $\beta$, shown in 
Fig.~\ref{Helicity}, tells us that, for a given $N$, $\eta$ in the BKT phase 
decreases monotonically from a value compatible with $\eta^{(1)}=1/4$, taken 
at the first critical coupling $\beta_c^{(1)}$, to a value compatible with 
$\eta^{(2)}=4/N^2$, taken at the second critical coupling. In some sense, the
$N^2$ drop of the value of $\eta$ at the second transition is related to
increasing distance between $\beta_c^{(2)}$ and $\beta_c^{(1)}$. 

We studied also the specific heat at the two transitions, finding that,
in contrast to the case of first- and second-order phase transitions, 
it does not reflect any nonanalytical critical properties at the critical 
temperatures, thus confirming that only BKT transitions are at work here.  

\section{Behavior with $N$ of the critical couplings}
\label{sec:N_dep}

The results of this work and those available in the literature allow us to
make some considerations about the behavior with $N$ of the critical coupling 
$\beta_c^{(1)}$ and $\beta_c^{(2)}$. 
Examining Table~\ref{tbl:betaci} one concludes that 
formulae~(\ref{betac1appr}) and~(\ref{betac2approx}) give the correct 
qualitative prediction for the scaling of the $\beta_c^{(1,2)}$ with $N$ in 
the Villain formulation. Namely,
\begin{itemize}
\item 
$\beta_c^{(1)}$ converges to $XY$ value very fast, like $\exp(-a N^2)$
\item 
$\beta_c^{(2)}$ diverges like $N^2$ \ .
\end{itemize}
One should expect that the standard vector $Z(N)$ model possesses similar 
scaling, probably up to ${\cal{O}}(N)$ corrections. One could try therefore 
to fit available Monte Carlo data for $\beta_c^{(1,2)}$ with 
formulae~(\ref{betac1appr}) and~(\ref{betac2approx}) modified 
to account for such corrections.     

\begin{figure}[tb]
\begin{center}
\includegraphics[scale=0.5]{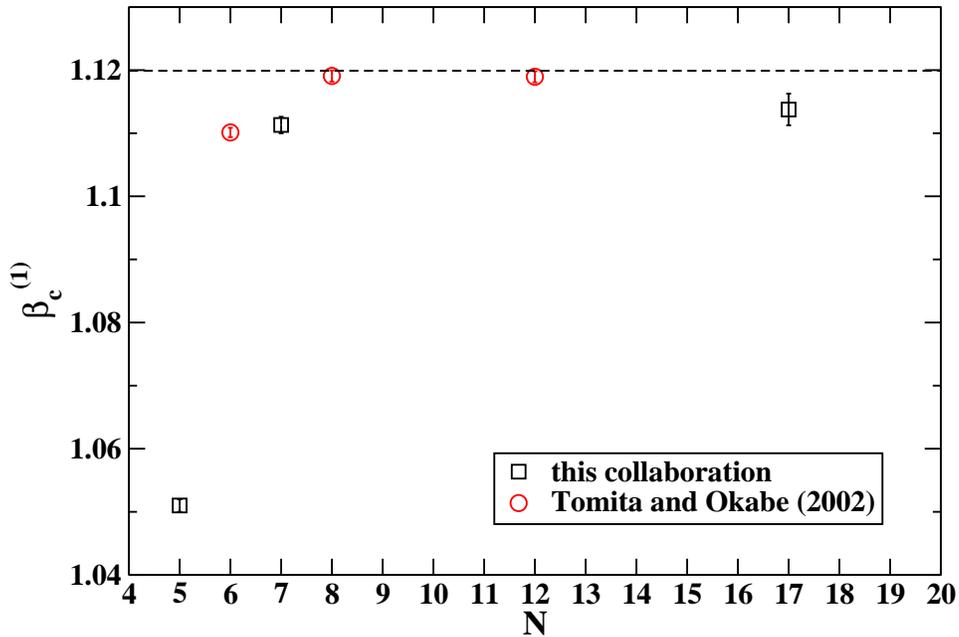}
\caption{(Color online) Behavior with $N$ of the known values of the critical 
coupling $\beta_c^{(1)}$ in $2D$ $Z(N)$ vector models. The horizontal dashed 
line represents the critical coupling of $2D$ $XY$, $\beta_{XY}$=1.1199, 
taken from Ref.~\cite{Hasenbusch:2005xm}.}
\label{crit_beta1}
\end{center}
\end{figure}

\begin{figure}[tb]
\begin{center}
\includegraphics[scale=0.5]{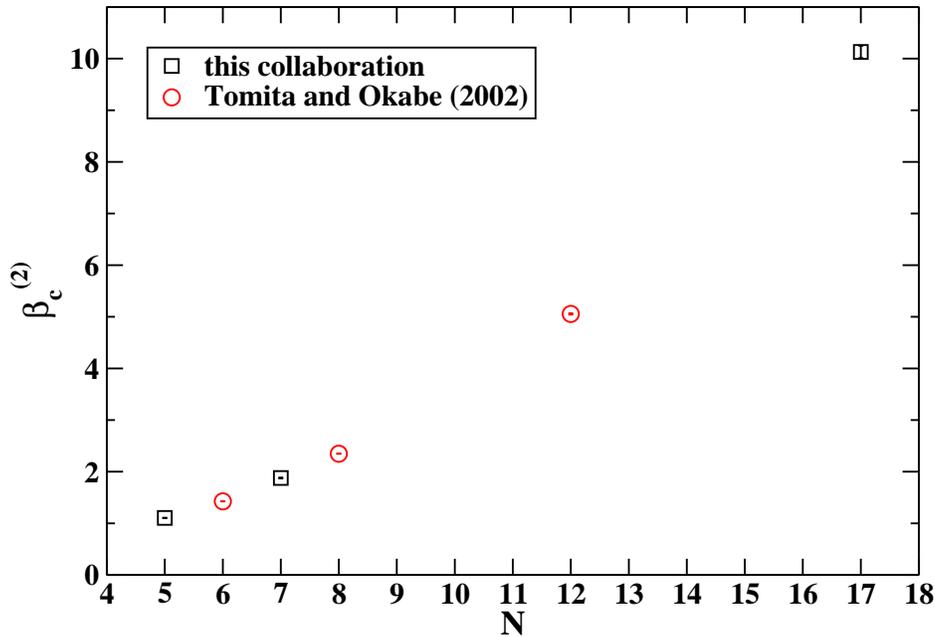}
\caption{(Color online) Behavior with $N$ of the known values of the critical 
coupling $\beta_c^{(2)}$ in $2D$ $Z(N)$ vector models.}
\label{crit_beta2}
\end{center}
\end{figure}

In Table~\ref{N_dep} and in Figs.~\ref{crit_beta1},~\ref{crit_beta2} we 
summarize the present knowledge about the position of the critical points for 
$2D$ $Z(N)$ vector models.

\begin{table}[tb]\setlength{\tabcolsep}{12pt}
\centering
\caption[]{Summary of the known values of the critical couplings 
$\beta_c^{(1)}$ and $\beta_c^{(2)}$ for $2D$ $Z(N)$ vector models.}
\vspace{0.2cm}
\begin{tabular}{|c|l|l|c|}
\hline
$N$ & \hspace{0.7cm} $\beta_c^{(1)}$ & \hspace{0.7cm} $\beta_c^{(2)}$ 
& Reference \\
\hline
 5  & 1.0510(10)      & \hspace{0.11cm} 1.1048(10)  & \cite{z5_phys.rev} \\
 6  & 1.11012(74)     & \hspace{0.11cm} 1.4257(22)  & \cite{cluster2d}    \\
 7  & 1.1113(13)      & \hspace{0.11cm} 1.8775(75)  & this work           \\
 8  & 1.11907(88)     & \hspace{0.11cm} 2.3480(22)  & \cite{cluster2d}    \\
12  & 1.11894(88)     & \hspace{0.11cm} 5.0556(128) & \cite{cluster2d}    \\
17  & 1.11375(250)    & 10.13(12)   & this work           \\ 
$\infty$ & 1.1199(1)  & \hspace{0.7cm} $\infty$   & \cite{Hasenbusch:2005xm} \\
\hline
\end{tabular}
\label{N_dep}
\end{table}

We can see from Fig.~\ref{crit_beta1} that the approach of $\beta_c^{(1)}$
to the $N\to\infty$ limit, corresponding to the $2D$ $XY$ model, is indeed 
very fast. Introducing corrections into the scaling formula~(\ref{betac1appr}),
we may conjecture the following general behavior:
\begin{equation}
\beta_c^{(1)} \ = \ A - (B N^2 + C N + D) 
\exp{ \left [ - \frac {N^2}{E} \right ] } \ .
\label{mc1fit}
\end{equation}
Indeed, attempts to fit data with this formula give values for $A$ and $E$ 
compatible with $\beta_c^{XY}$. However, as is seen from Table~\ref{N_dep}, 
the combination of our Monte Carlo determinations and those of 
Ref.~\cite{cluster2d} leads to a fake non-monotonic approach to 
$\beta_c^{XY}$, in marked contradiction with our RG analysis. Looking
separately at the two data sets given in Table~\ref{N_dep}, one can see
that each set indeed satisfies monotonicity (taking into account the 
error bars in the case of data from Ref.~\cite{cluster2d}).
We think, therefore, that the non-monotonicity of the combined set can be 
explained with the different systematics affecting the determinations
of $\beta_c^{(1)}$ by the two collaborations. On our side, we have determined 
the location of $\beta_c^{(1)}$ by two different methods, using several 
observables and working on larger lattices, up to $L=1024$, thus making
us very confident on the reliability of our results. In conclusion,
a reliable check of the formula~(\ref{mc1fit}), based on all data given 
in Table~\ref{N_dep}, is not possible and that formula remains a conjecture.

In the case of $\beta_c^{(2)}$, instead, one can verify that, for example,
the following extension of~(\ref{betac2approx})
\begin{equation} 
\beta_c^{(2)} \ = \ \frac{N^2}{A} + B N + C + D e^{- \frac{\pi^2 N^2}{A}} 
\label{mc2fit}
\end{equation}
fits the data rather well, although with a high $\chi^2$/d.o.f.
probably reflecting the different systematics mentioned above,
$$
A\ = \ 25.89(115) \ , \;\;\; B \ = \ -0.040(23) \ , \;\;\; C \ = \ 0.38(9) \ , 
\;\;\; D \ = \ 227.7(634) \ , \;\;\; \chi^2 \ = \ 12.69 \ .
$$
The value of $A$ is close to $8\pi$, as is expected. 

In general, however, we have to conclude that even if the scaling 
formulae~(\ref{mc1fit}) and~(\ref{mc2fit}) might be correct, more data for 
different $N$ are needed to determine all coefficients.   

\section{Summary}

In this paper we have studied the RG equations describing the critical 
behavior of $2D$ $Z(N)$ vector models in the Villain formulation. 
The main original results are
\begin{itemize}
\item the RG trajectories in the vicinity of both phase transitions, 
\item the critical points as functions of $N$,
\item the index $\nu$, which turns out to be equal to 1/2 for all $N>4$. 
\end{itemize}

The numerical part of the work has been devoted to verify these
theoretical expectations:
\begin{itemize}
\item We have determined numerically the two critical couplings of the $2D$
$Z(N=7, 17)$ vector models and given estimates of the critical indices $\eta$
at both transitions. Our findings support for all $N\geq 5$ the standard 
scenario of three phases: a disordered phase at high temperatures, a
massless or BKT one at intermediate temperatures and an ordered phase, 
occurring at lower and lower temperatures as $N$ increases. 
This matches perfectly with the $N\to\infty$ limit, {\it i.e.} the $2D$ $XY$ 
model, where the ordered phase is absent or, equivalently, appears at 
$\beta\to\infty$. 
\item We have found that the values of the critical index $\eta$ at the two 
transitions are compatible with the theoretical expectations. 
\item The index $\nu$ also appears to be compatible with the value $1/2$, 
in agreement with RG predictions. 
\end{itemize}

On the basis of this study and taking into account previous 
works~\cite{cluster2d,z5_phys.rev}, we are prompted to conclude that 
$2D$ $Z(N)$ vector models in the standard formulation undergo two phase 
transitions of the BKT type. Furthermore, the standard and the Villain 
formulations belong to the same universality class with $\nu=1/2$, 
$\eta^{(1)}=1/4$ and $\eta^{(2)}=4/N^2$.

Considering the determinations of the critical couplings as a function of $N$, 
we have calculated the leading dependences using RG equations and conjectured 
the approximate scaling for $\beta_c^{(1,2)}(N)$ in the standard version. 
The existing numerical values and their accuracy are not sufficient, however 
to reliably check the conjectured formulae.  

\section{Acknowledgments}

The work of G.C. and M.G. was supported in part by the European Union 
un\-der ITN STRO\-NG\-net (grant PITN-GA-2009-238353).

\end{document}